\documentstyle[11pt,paspconf,psfig]{article}

\begin{document}

\title{Galaxy Formation and Large Scale Structure}

\author{Richard S. Ellis}
\affil{Institute of Astronomy, University of Cambridge
    CB3 0HA, England}

\begin{abstract}

Galaxies represent the visible fabric of the Universe and there
has been considerable progress recently in both observational and
theoretical studies. The underlying goal is to understand the
present-day diversity of galaxy forms, masses and luminosities in
the context of theories for the growth of structure. Popular
models predict the bulk of the galaxy population assembled
recently, in apparent agreement with optical and near-infrared
observations. However, detailed conclusions rely crucially on the
choice of the cosmological parameters. Although the star formation
history has been sketched to early times, uncertainties remain,
particularly in connecting to the underlying mass assembly rate. I
discuss the expected progress in determining the cosmological
parameters and address the question of which observations would
most accurately check contemporary models for the origin of the
Hubble sequence. The new generation of ground-based and future
space-based large telescopes, equipped with instrumentation
appropriate for studying the resolved properties of distant
galaxies, offer the potential of substantial progress in this
field.

\end{abstract}


\section{Introduction}

Despite a dramatic surge of interest in observational cosmology
(Figure~\ref{fig-1}), our physical understanding of how the local
diverse population of galaxies formed remains unclear. Although
semi-analytical models based on hierarchical growth in a cold dark
matter dominated Universe are able to reproduce many basic
observables such as the joint distributions of magnitude, colour
and redshift to faint limits, and the spatial distribution of
sources selected in different ways at various look-back times, we
seek more fundamental ways of making progress and, in particular,
of finding observational tests of galaxy formation models immune
from uncertainties introduced by our poor knowledge of the
cosmological parameters and the power spectrum of mass
fluctuations. In this brief non-specialist review, I assess the
renewed enthusiasm for independently determining the cosmological
parameters and emphasise how the new generation of ground-based
and future space-based large telescopes, equipped with
instrumentation appropriate for studying the resolved properties
of distant galaxies, might then approach the question of
understanding the origin of the Hubble sequence of morphological
types.

\begin{figure}
\caption{The
rapid growth in publications in Ap J and AJ containing the keyword
{\it observational cosmology}. The dates of some significant
events are included. Keyword statistics cannot be reliably used
prior to 1975.} \label{fig-1}
\end{figure}

\section{Models for the Growth of Structure}

The most popular model for the origin and development of structure
in the expanding Universe is the cold dark matter (CDM) model
(White \& Rees 1978, Blumenthal et al 1984) whose predictive power
for both the assembly of mass structures and the history of galaxy
formation has been exploited via both numerical simulations (Frenk
et al 1985) and semi-analytical calculations (Kauffmann et al
1993, Cole et al 1994, 1999).

CDM posits that the bulk of gravitating mass in the Universe is
non-baryonic in form and which detaches from the expanding plasma
prior to recombination. A present-day galaxy can be regarded as a
peak of radiating baryonic matter embedded within a more extensive
dark matter halo. Although the growth of mass structures is
governed by the merging hierarchy of dark matter halos (Press \&
Schechter 1985), the onset of galaxy formation depends on the
interplay between gas cooling and star formation which relies on
more detailed astrophysics. Further development e.g. details of
the formation of the Hubble sequence remains unclear. CDM should
therefore be envisioned as a theory of structure formation
underpinned by a fundamental assumption - the presence of copious
amount of non-interacting dark matter - upon which more detailed
astrophysical ingredients must be added.

Even in the context of understanding large scale structure, the
CDM picture represents a working framework rather than a specific
well-defined model. Within hierarchical dark-matter dominated
models are variants which differ in the assumed cosmological
parameters ($\Omega_{mass}$, $\Omega_{baryon}$,
$\Omega_{\Lambda}$, $H_o$ etc) and the spectrum of initial
fluctuations, characterised by the slope $n$ ( $n$=1 in most
inflationary models) and absolute normalisation (the latter quoted
in terms of the variance of mass fluctuations on 8$h^{-1}$ Mpc
scales, $\sigma_8$). None of these key parameters is known with
certainty, including perhaps even $H_o$, but three combined sets consistent
with much of the available data are in popular use (see
Table \ref{table-1}) (adapted from Thomas et al 1998). Each
represents a variant of the original 'Standard' SCDM which fails
to jointly fit the absolute value of microwave background
fluctuations found by COBE and the present-day number density of
rich clusters (Wright et al 1992).

\begin{table}
\caption{The CDM Family}
\label{table-1}
\begin{center}
\begin{tabular}{crrrrrr}
CDM Variant & $\Omega_{mass}$ & $\Omega_{baryon}$ &
$\Omega_{\Lambda}$ & $H_0$ & $n$ & $\sigma_8$ \\ \\
\tableline

Standard SCDM & 1.0 & 0.050 & 0.0 & 50 & 1 & 0.61 \\

\tableline

Open OCDM & 0.3 & 0.026 & 0.0 & 70 & 1.3 & 0.85 \\

Lambda $\Lambda$CDM & 0.3 & 0.026 & 0.7 & 70 & 1 & 1.30 \\

Tilted $\tau$CDM   & 1.0 & 0.050 & 0.0 & 50 & 0.7 & 0.68 \\

\end{tabular}
\end{center}
\end{table}

The key observables which have, over the past few years, been used
to reject SCDM and support some or all of the others include the
absolute normalisation of large scale fluctuations in the
microwave background (Bunn \& White 1997), the local number
density of rich clusters of galaxies (White et al 1993) and the
power spectrum of the local galaxy distribution (Peacock 1997).
Broadly speaking, the CDM variants are underconstrained by these
observables if the cosmological parameters are allowed a free
range. However, structure {\it evolution} over 0$<z<$2, during
which time the local galaxy population might have assembled,
depends sensitively on the values of $\Omega_M$ and
$\Omega_{\Lambda}$ and thus if it became possible to independently
constrain these parameters, systematic observations of distant
galaxies would provide less ambiguous information on physical
aspects of galaxy formation. The bottom line is that it will be
hard to find a shortcut to substantial progress in understanding galaxy
formation without comparable effort to nailing down the cosmological 
parameters.

\section{Progress in Structure Formation and Cosmology}

Should we believe the growing optimism that we will soon resolve
many of the key variables that govern the history of large scale
structure, including the elusive cosmological parameters (e.g.
Tytler 1997, Turner 1999) - or are we witnessing a rerun of the
enthusiasm that swept observational astronomy in the 1970's (e.g.
Gunn \& Tinsley 1975, Gott et al 1976)? I have selected four basic
observational initiatives and good progress can be expected for
each of them in the next decade.

\subsection{The Local Distribution of Galaxies}

The local spectrum of mass fluctuations has, until now, been
largely constrained by the abundance of rich clusters quantified
via $\sigma_8$ (Eke et al 1996) (see Table~\ref{table-1}).
However, clusters of galaxies cannot be easily selected on an
uniform basis in terms of their total mass. Cluster mass estimates
based on galaxy dynamics are affected by substructures whose
nature is poorly defined even when hundreds of velocities are
available. Likewise, masses based on X-ray luminosities or
temperatures rely on structural data which is in short supply. The
power spectrum $P(k)$ \footnote{$k=2\pi/\lambda$ is the Fourier
wavenumber of a particular comoving length scale, $\lambda$ Mpc.}
of a large uniformly-selected galaxy redshift survey constrains
the mass distribution on large scales with the disadvantage that
galaxies may be biased tracers of the underlying matter. When
coupled with the amplitude of angular fluctuations in the
microwave background, joint constraints on various cosmological
parameters become available (see Eisenstein et al 1998).

Very large, statistically-complete, redshift surveys are now
underway (Gunn \& Weinberg 1995, Colless 1999) motivated by the
need to constrain the power spectrum and topology of the galaxy
distribution as well as the peculiar velocity field, either
statistically or through specific surveys targetting individual
classes of galaxies (Colless et al 1999). At the time of writing
the largest local survey is the Anglo-Australian 2dF Galaxy
Redshift Survey (2dFGRS -- Taylor 1995, Colless 1999) which has
$\simeq$50,000 redshifts to $b_J$=19.5 from a currently-incomplete
magnitude-limited sample within one 75$^{\circ}\times$5-15
declination strip in each Galactic hemisphere. An important
development with these surveys is the generation of 'mock
catalogues' (Cole et al 1998) based on numerical simulations
incorporating bias prescriptions, observational selection criteria
and the instrumental characteristics of the actual surveys
(Figure~\ref{fig-2}).

\begin{figure}
\caption{(Top) The redshift-space distribution of galaxies so far
in the Anglo-Australian 2dF Galaxy Redshift Survey (Colless 1999).
(Bottom) That expected in $\tau$CDM from the mock catalogue
produced by Cole et al (1998). The mock catalogue incorporates
many of the technical limitations of the 2dF instrument and
details of the survey strategy as well as the manner in which
galaxies are biased.}
\label{fig-2}
\end{figure}

Deeper surveys can, in principle, constrain the correlation
function of galaxies at large look-back times, in order to track
the evolution of structure. In practice however, there are
formidable obstacles. Surveys such as 2dFGRS are magnitude-limited
and, faintward, k-corrections increase the spread in redshift
considerably so that the mean depth $\overline{z}$ is not strongly
dependent on magnitude limit (Ellis et al 1996). A $B$=25
wide-field 8-metre telescope survey is likely to have a
significant volume overlap with one surveyed to $B$=22 by 2dF
itself (although the galaxies sampled would of course be
intrinsically less luminous).

If galaxies can be selected to lie in specific redshift intervals
through judicious photometric selection (e.g. the Lyman dropout
technique) then {\it angular} correlations can be more effective
(Steidel et al 1996). The difficulty is relating restricted
populations to those observed locally. However, this problem
remains in pure magnitude-limited samples since k-corrections
distort the observed mix of types each of which may have different
spatial distributions. A number of groups are now undertaking
multi-band panoramic surveys with a view to utilising photometric
techniques to examine angular correlations as a function of
inferred redshift (Brunner et al 1999, Marzke 1999,
Figure~\ref{fig-3}). Such analyses will necessitate complex
modelling to account for the subtle selection effects involved
(Kauffmann et al 1999).

In biased galaxy formation, early star-forming systems, e.g. Lyman
break galaxies, should be more strongly clustered (or biased) than
their later versions forming at lower redshift. Bias can be tested
within a given dataset by correlating the clustering amplitude
with luminosity or preferably mass. Luminosity-dependent
clustering is tentatively claimed for the Lyman break galaxies
(Steidel et al 1999) and interpretation of this signal will depend
on deriving the dynamical masses of individual examples (Pettini
et al 1999).

\begin{figure}
\caption{A new generation of faint panoramic optical and infrared
imaging surveys is motivated by the desire to constrain evolution in
the angular clustering of galaxies as a function of epoch. The image
represents a 13 $\times$13 arcmin$^2$ field limited taken
with four mosaiced exposures using CIRSI - a 4 $\times$ 1024$^2$ HgCdTe
array (Beckett et al 1998).  Analyses of angular correlations for
IR-selected samples partitioned according to photometric redshift will
necessitate consideration of both k-correction and luminosity-dependent
biases.}
\label{fig-3}
\end{figure}

In the next few years we can thus expect a huge influx of z$<$0.3
redshift survey data, the completion of several deep panoramic
optical-infrared imaging surveys and detailed studies of selective
populations of galaxies at z$>$1. These data will be influential
in understanding biased galaxy formation and constraining the
power spectrum $P(k)$ and recent growth in clustering.

\subsection{Angular Fluctuations in the Microwave Background}

Proponents of the two major microwave background satellites of the
next decade (the Microwave Anistropy Probe (MAP, {\tt
http://map.gsfc.nasa.gov}) and particularly the Planck Surveyor
{\tt http://astro.estec.esa.nl/Planck/}) promise precision
measurement of all cosmological parameters and characteristics of
the initial power spectrum. Is such optimism simply a reflection
of the fact that the theoretical modelling is well ahead of the
observational data?

A good assessment of the immediate opportunities can be found by
considering evidence for the location of a possible `Dopper peak'
in the spectrum of angular fluctuations delineated by various
experiments (Figure~\ref{fig-4}). The sought-after feature is the
sound horizon at recombination and its angular scale measures the
angular diameter distance at $z\simeq$1100 and hence the spatial
curvature representing the sum of the contributing energy
densities $\Omega_M\,+\,\Omega_{\Lambda}$. Efstathiou et al (1999)
\footnote{See also the independent analyses by Lineweaver (1998)
and Tegmark (1998).} analysed the extant data and found:

$$\Omega_M\,+\,\Omega_{\Lambda} \equiv 0.88 \pm 0.06 (1\sigma) $$

\begin{figure}
\caption{Angular spectrum of fluctuations in
the microwave background as constrained by various experiments
(see legend) leading to the suggestion of a `Doppler peak' whose
angular scale constrains the spatial curvature
$\Omega_M\,+\,\Omega_{\Lambda}$ to be flat (after Efstathiou et al
1999)} \label{fig-4}
\end{figure}

Considerable progress is expected in the next few years in
confirming or otherwise this result which, at the moment, involves
combining measurements made with different instruments. The
Boomerang project\footnote{{\tt
http://astro.caltech.edu/~lgg/boom/boom.html}} should be the first
to trace the fluctuation spectrum through the putative Doppler
peak with a single instrument.

\subsection{Supernovae and The Cosmic Deceleration}

The traditional apparent magnitude redshift relation for brightest
cluster galaxies led to conflicting claims on the cosmic
deceleration parameter $q_0$ \footnote{The deceleration parameter
$q_0= -\ddot{R}\,R/\dot{R}^2$, where $R(t)$ is the scale factor,
is no longer commonly used as it maps onto $\Omega_M$ and
$\Omega_{\Lambda}$ in a redshift-dependent manner.}(Gunn \& Oke
1974, Kristian et al 1977). Uncertain corrections for luminosity
evolution in giant ellipticals even led to early claims for a
cosmic acceleration (Gunn \& Tinsley 1975)! A more recent infrared
study (Arag\'on-Salamanca et al 1998) introduced a further
correction for the assembly of stellar mass from mergers since
$z\simeq$1. First-ranked cluster members are evidently complex
evolving systems whose properties will have to be understood in
considerable detail before any believable constraints on the
cosmic deceleration emerge.

It is thus only reasonable to be skeptical about recent claims
(Garnavich et al 1998, Perlmutter et al 1999) for a cosmic
acceleration based on the apparent magnitude - redshift relation
for distant Type Ia supernovae (SNIa, (Figure~\ref{fig-5}). May
their poorly-understand progenitors not evolve in some as yet
undiscovered way with redshift?

\begin{figure}
\caption{The Hubble diagram (top) for Type Ia
supernovae published by the Supernova Cosmology Project
(Perlmutter et al 1999). The dispersion in rest-frame blue
apparent magnitude, corrected for light curve width - peak
luminosity correlations, remains remarkably tight at all redshifts
(inset). Distant supernovae are $\simeq$0.5 mag too faint to be
consistent with a spatially-flat matter-dominated Universe
(bottom) ([$\Omega_M:\Omega_{\Lambda}$]=[1,0]) and provide
tantalising evidence for a non-zero $\Omega_{\Lambda}$. Individual
outliers (unfilled circles) with incomplete or unsatisfactory data
do not affect this conclusion.} \label{fig-5}
\end{figure}

There are, however, fundamental differences between the
first-ranked cluster galaxy and SNIa campaigns. Foremost,
supernovae are {\it thermodynamical events} not a restricted class
of objects drawn in a (possibly biased) manner from a wider
statistical population. Secondly, the information obtained for
each distant SN event is impressive and includes (i)
time-dependent spectra for day-by-day comparison with that of
local examples constraining possible compositional and energetic
differences, (ii) light curves in various photometric bands useful
for clarifying correlations between the rest-frame peak luminosity
and light curve shapes, and (iii) the morphological properties of
the host galaxies and the location of the SN event within them.
With larger samples, analyses can be executed for subsamples
representing, e.g. the most luminous examples, those found in
dust-free environments or with old stellar populations. The
absence of an increased scatter in Figure~\ref{fig-5} at large
redshift is highly significant. If progenitor evolution, host
galaxy dust or other systematic effects were distorting the curve
towards a non-zero $\Lambda$, then such effects would have be
orchestrated to be similar across a wide range of environments and
galaxy evolutionary histories.

The SNIa data do not yet span a sufficiently large redshift range
to independently constrain $\Omega_M$ and $\Omega_{\Lambda}$ and
thus measures deceleration over a particular redshift range,
effectively $\Omega_M\,-\,\Omega_{\Lambda}$. Contrary to popular
perception, the strongest result is the rejection of the
matter-dominated Einstein-de Sitter model {\it not} the presence
of a non-zero cosmological constant (the latter being only a
3$\sigma$ effect with the current data). Only by combining the SN
constraint with that of spatial curvature derived from the
location of the microwave background Doppler peak is strong
evidence for a cosmic acceleration inferred (Figure~\ref{fig-6},
Efstathiou et al 1999).

\begin{figure}
\caption{The location of the Doppler peak in the microwave background
(which constrains $\Omega_M\,+\,\Omega_{\Lambda}$ - Figure~\ref{fig-4})
can be combined with complementary constraints on
$\Omega_M\,-\,\Omega_{\Lambda}$ from the Type Ia supernova data
(Figure~\ref{fig-5}) (Efstathiou et al 1999). The resulting solution
supports a spatially flat
($\Omega_K$=1\,-\,($\Omega_M\,+\,\Omega_{\Lambda}$)=0) Universe with
$\Omega_M$=0.3, $\Omega_{\Lambda}$=0.7. }
\label{fig-6}
\end{figure}

What are the weak points in the SNe analyses? Notwithstanding the
near-constant scatter with redshift referred to above, uniformly
distributed dust extinction (arising within all host galaxies at
$z>$0.5 or in the intergalactic medium - Aguirre 1999) and
progenitor evolution (c.f. Riess et al 1999) remain a concern.
Fortunately, we can expect to address both convincingly in the
next few years via much larger samples from which
carefully-selected subsets can be compiled, and by extending the
analysis beyond z$\simeq$1 where the effects of a non-zero
$\Lambda$ should {\it decrease} in comparison with that of a
uniform absorbing medium. The challenge, as always, lies in
designing the most effective facility to conduct the necessary
survey. Considering the impact of the SNe Ia results, it is surely
high time to consider more ambitious instrumentation dedicated to
locating them and their multi-band follow-up.

\section{Gravitational Lensing and the Mean Mass Density}

The synergy achieved between the SNe and microwave background data
is, at first sight, reassuring (Ostriker \& Steinhardt 1995).
However, the conclusions lead cosmology into difficult territory.
What is the physical origin of a non-zero $\Lambda$ and why is its
energy density anywhere near that of the gravitating mass?
(Steinhardt et al 1999)?

Not only do we seek verification via at least one further
constraint on Figure~\ref{fig-6} but also, in understanding galaxy
formation, the relative distribution of baryonic and non-baryonic
matter and precise constraints on $\Omega_M$ are important goals.
The most fundamental assumption of CDM is the presence of
non-baryonic matter and, short of its direct detection, comparing
the baryonic and non-baryonic components is an essential test.
Mass distributions on cosmological scales can be constrained with
ambitious wide field imaging surveys utilising the weak
gravitational lensing of background field galaxies (Narayan \&
Bartelmann 1996, Mellier 1998, ).

Gravitational lensing is a fascinating phenomenon. However, even
its most ardent supporters would admit it has yet to provide
convincing constraints on the distribution of non-baryonic matter
on scales upward of 1 Mpc. Work has concentrated where the signals
are strongest, i.e. the cores of rich clusters. Here weak shear
and strong lensing (multiply-imaged arcs) studies have confirmed
high M/L ratios consistent with $\Omega_M\simeq$0.3 inferred
already from dynamical and X-ray studies (Smail et al 1997, Allen
1998).

Powerful techniques have now been developed to derive the
projected distribution of lensing mass from the weak shear seen
across the field of a CCD detector. The methods assume that,
statistically, the intrinsic ellipticity and redshift distribution
of the background sources are well-behaved (Kaiser \& Squires
1996, Kaiser et al 1998, Schneider et al 1998). In practice, the
techniques are limited by instrumental and seeing corrections,
particularly for ground-based data. Comparative studies utilising
different techniques for the same rich cluster observed
independently with different telescopes (Erben et al 1999) suggest
mass distributions can be extracted reliably to sensitivity levels
equivalent to those needed to see coherent shear from larger-scale
structures {\it outside} rich clusters. Potential targets for weak
lensing studies therefore now extend to individual filaments whose
existence is inferred from redshift surveys or otherwise (e.g.
Kaiser et al 1998, Figure~\ref{fig-7}).

\begin{figure}
\caption{The extension of weak gravitational lensing as a probe of 
mass distributions outside rich clusters is an imminent prospect with 
panoramic imaging cameras. The image shows Kaiser et al's (1999) mass
reconstruction in a field containing two X-ray clusters. Dark material
connecting such systems may be revealed through weak lensing techniques.}

\label{fig-7}
\end{figure}

There is great potential because lensing offers a bias-free probe
of mass fluctuations on various scales (Jain \& Seljak 1997). CCD
cameras with upwards of 8000$^2$ pixels make surveying for mass
structures independently of their baryonic fraction a practical
prospect provided the systematics can be controlled. Correlation
studies of the shear drawn from many randomly-chosen independent
fields can provide a global measure of the mass power spectrum on
a particular scale (the `cosmic shear' - Mould et al 1994) whereas
ultimately one can hope to define projected mass maps and
interpret these statistically for $\Omega_M$ by correlating with
similar data from redshift surveys (Figure~\ref{fig-2}).

In conclusion, there are four observational programmes each of
which already offers complementary constraints on the cosmological
parameters and power spectrum of mass fluctuations. More
importantly, the future prospects for each are excellent. By
eliminating the cosmological uncertainties which underpin all
theories for structure formation, we can hope to move forwards to
a physical picture of how galaxies form and evolve.

\section{The Origin of the Hubble Sequence}

Although originally introduced for taxonomic purposes, the Hubble
sequence of morphological types defines an axis along which there
are strong physical trends. For stellar populations the sequence
represents the locus of the ratio of the current star formation
rate to a long-term average (Struck-Marcell \& Tinsley 1978).
There are near-monotonic dependences in gas content, rotational
support, bulge/disk ratio and central density (de Vaucouleurs
1977, Efstathiou \& Silk 1983). Most importantly, there are
suggestive variations in the population mix as a function of
environment (Dressler 1980, Whitmore et al 1993).

The challenge to infer the evolutionary history of this sequence
has been revolutionised by Hubble Space Telescope's (HST) ability
to resolve kpc scale structures at most redshifts of interest.
However, surprisingly little progress has been made in equivalent
2-D spectroscopic follow-up of these distant sources with
ground-based telescopes. This is one of the key areas of
importance in the next decade.

\subsection{Monolithic and Hierarchical Theories of Galaxy Formation}

The continuous merging of dark matter halos and delayed onset of
star formation in the hierarchical picture has led to a profound
challenge to classical views for the origin of the Hubble
sequence. Prior to CDM, the traditional viewpoint (e.g. Eggen et
al 1962, Sandage 1986) associated the high stellar density, low
specific angular momentum and homogeneous old stellar populations
of spheroidal galaxies with rapid dissipationless collapse at high
redshift. Dissipative collapse of gas clouds at later epochs would
lead to rotationally-supported disks destined to become local
spirals. Giant ellipticals forming at high redshift via
`monolithic' collapse prompted numerous searches for luminous
primaeval galaxies (Djorgovski 1996).

In hierarchical pictures however, the first stellar systems will
be small and numerous, reflecting the abundance of
appropriately-sized halos at early times. Rapid merging of these
initial systems delivers a population of young bulges. Continued
gas cooling produces stable disks around these bulges some of
which will later merge to form large spheroidal galaxies (Baugh et
al 1995).

The differences between the monolithic and hierarchical pictures
are greatest for the spheroidal galaxies. Merger-driven
morphological evolution occurs earlier in the denser environments
(Kauffman 1995) and thus strong differential effects are also
expected in the sense that field spheroidals will always be
younger and more inhomogeneous than their clustered equivalents.
Although evidence continues to accumulate suggesting that many
local ellipticals formed via the merger of disk galaxies
(Schweizer 1997), this does not necessarily imply {\it all}
spheroidals formed recently.

\subsection{Cosmic Star Formation and Mass Assembly Histories}

The completion of extensive redshift surveys (Lilly et al 1995,
Ellis et al 1996, Cowie et al 1999) and the identification of
representative samples of Lyman break galaxies at various epochs
(Steidel et al 1996, 1999) has led to analyses of the comoving
density of star formation as a function of look-back time (Madau
et al 1996, Madau 1999, Figure~\ref{fig-8}). Does this 'Madau
plot' which indicates an extended history of star formation
differentiate between hierarchical and monolithic galaxy formation
as has been claimed (Baugh et al 1998)?

\begin{figure}
\caption{The redshift-dependent comoving volume-averaged density of
star formation derived from the luminosity density of ultraviolet
light, H$\alpha$-emission and bolometric far-infrared flux (Madau
1999). Data points with error bars refer to UV-based measures with no
extinction correction (left panel) and that corrected with a constant
value at 1500\AA\ (right panel).  The dotted line shows the fiducial
value necessary to produce the observed background light. No star
formation diagnostic is immune from difficulties and comparisons for
the same sources suggest further complications arise from the
timescales over which a given diagnostic applies. }
\label{fig-8}
\end{figure}

In quantitative detail, the star formation history is susceptible
to major changes through unaccounted obscured sources, dust
extinction on the detected ones, incompleteness arising from
sources fainter than the survey limit, and inaccuracies or
inappropriate comparisons associated with different diagnostics
used, each of which has its own limitations. Although in broad
agreement with the CDM predictions (Baugh et al 1998), detailed
comparisons depend more on how feedback and star formation are
implemented in the semi-analytical models rather than on structure
formation in a cold dark matter Universe (c.f. Jimenez et al
1999). Although Figure~\ref{fig-8} has been central in motivating
optimum strategies for finding distant sources, we now seek a more
fundamental way to address the physical foundations of galaxy
formation theories. Ultimately this implies tracking the evolution
of the assembling mass.

\subsection{Field Spheroidals}

The continuous formation of field spheroidals from merging spirals
contrasts markedly with the traditional picture of monolithic
collapse at high redshift. This seems one of the simplest ways to
differentiate two very different pictures of morphological
evolution. However, in practice, a demise with redshift in the
abundance of field ellipticals has been remarkably difficult to
verify. Although field ellipticals can be readily discerned to
$I$=22-23 in typical WFPC-2 images (and $I$=24-25 in the higher
signal to noise Hubble Deep Fields), morphological
number-magnitude counts (Glazebrook et al 1995, Driver et al 1996,
Abraham et al 1997) give ambiguous results because of
uncertainties associated with the local luminosity function.
Regardless, the evolution expected in hierarchical CDM is a strong
function of the cosmological model, with little evolution to
$z\simeq$1-2 expected in open or $\Lambda$-dominated versions
(Kauffmann \& Charlot 1998b).

Optical-infrared colours offer an alternative route to
constraining the assembly history. In the monolithic case we would
expect to find a sizeable population of intrinsically red systems
beyond z$\simeq$1; in fact few are found (Zepf 1997, Barger et al
1999). Menanteau et al (1999) selected a large sample of
morphological spheroidals and demonstrated a marked paucity of
examples with colours compatible with passively evolving systems
formed at $z>$3. The drawback is that only a modest amount of
ongoing star formation occurring in objects which collapsed at
higher redshift would be needed to move passively-red systems into
the colour range observed (Jimenez et al 1999).

Ideally we seek a characteristic that demonstrates that field
ellipticals are bluer and more inhomogeneous than their clustered
counterparts from which the associated mass assembly rate could be
derived. Figure~\ref{fig-9} shows the internal colour distribution
of two rather different field spheroidals located in the HDF
(Abraham et al 1999a). Across a large sample of such systems,
evidence is accumulating that field ellipticals are more
inhomogeneous in their observed properties than those found in
clusters in qualitative agreement with the predictions of
hierarchical models. The patchy colour distribution of a
significant fraction of field spheroidals at z$>$0.3 may be
consistent with merger-driven evolution and the challenge is now
to quantify this in terms of both the observed merger fraction and
the mass assembly history. Inevitably, more detail is needed on
the metallicity and timescales of secondary star formation implied
in the HST images.

\begin{figure}
\caption {The internal V-I colour distribution
for two spheroidals of known redshift in the Hubble Deep Field
(Abraham et al 1999a). Such analyses indicate a greater degree of
internal inhomogeneity, possibly consistent with secondary star
formation induced by mergers, for field ellipticals than for their
clustered equivalents.}

\label{fig-9}
\end{figure}

\subsection{Faint Blue Irregulars}

Deep HST images confirmed that the bulk of the faint blue excess
first located in faint galaxy counts arises from galaxies of
irregular morphology (Glazebrook et al 1995, Ellis 1997). Such
systems also contribute significantly to the increasing star
formation density to z$\simeq$1 (Brinchmann et al 1998). But what
is the physical origin of these systems and why are so few found
today?

Pixel-by-pixel simulations which attempt to recover the appearance
of local galaxies as viewed at various redshifts taking into
account instrumental and redshift-dependent effects (Abraham et al
1998, Brinchmann et al 1998) suggest that only a small fraction of
the HST irregulars might be late-type spirals viewed at rest-frame
UV wavelengths (a claim now supported by the NICMOS images,
Figure~\ref{fig-10}).

\begin{figure}
\caption {A morphologically irregular galaxy in the HDF with the 
integral field unit for the Gemini CIRPASS spectrograph superimposed.
Each fiber would sample a 0.15 arcsec diameter portion of the galaxy
enabling detailed internal studies.}
\label{fig-10}
\end{figure}

The demise of these abundant peculiar systems with active star
formation remains an outstanding question. High resolution
infrared images are needed to quantify the established stellar
mass (Broadhurst et al 1992, Kauffmann \& Charlot 1998a) as well
as resolved dynamical data to address the possibility that many
merge to form more regular systems.

This will be an enormous challenge. Merging statistics based on
projected imaging data (Patton et al 1998, LeFevre et al 1999) are
difficult to interpret in the context of hierarchical models
because merger timescales are needed to convert pair fractions
into merger rates. Will spectroscopy help in resolving this
impasse? Possibly not in individual cases such as
Figure~\ref{fig-10}, but the redshift-dependent mass distribution
and star formation timescales of young peculiars will be an
important step forward in understanding the role of feedback which
apparently delays the formation of these systems.

\subsection{Disks, Bars and Bulges}

Numerical simulations have yet to make much headway in addressing
the quantitative evolution of {\it internal} features in galaxies,
e.g. stellar disks, bars and bulges, yet we can expect exquisite
observational data on these to z$\simeq$1 and beyond in coming
years. Theorists prefer to delay disk formation to late epochs in
order to avoid spiral destruction through mergers as well as an
inexorable transfer of angular momentum from the stellar disk to
the outer dark matter halo (Weil et al 1998).

Rotation curves are now available for almost a hundred distant
spirals to z$\simeq$1 (Vogt et al 1999) yet the associated
Tully-Fisher relation shows little sign of any evolution.
Structural decomposition of HST images indicate both increased
disk surface brightnesses and bluer colours consistent with a
higher star formation rate (Lilly et al 1998). All available data
are consistent with the notion that spirals had their current
space density and structural forms established at $z\simeq$1. Is
this in conflict with theory? Again, in open or
$\Lambda$-dominated models, the conflict might not be so severe.
Moreover, observations could be reconciled with predictions if
disks were still growing whilst their star formation rates
declined (Mo et al 1998). However, neither the observational data
nor the model predictions are sufficiently refined to be sure what
is happening at the moment.

Stellar bars and bulges offer a further clue to the growth history
of spiral galaxies. CDM models predict bulges should be the very
oldest components having formed at early times from mergers of the
first collapsing systems and therefore have intrinsically red
colours (Cole et al 1999). However, bulges selected from a sample
of HDF spirals reveal a remarkable range of colours with few as
red as ellipticals of the same redshift (Ellis \& Abraham 1999,
Figure~\ref{fig-11}) suggesting either many formed more recently
or a continous infall of gas or satellites introduced recent star
formation in their cores. Bulges may also form sequentially from
stellar bars, both features representing stages of instability in
a well-established disk embedded in a significant stable dark
matter halo (Combes 1999). A paucity of barred systems has been
claimed in the HDF (Abraham et al 1999b) although the sample size
remains small.

\begin{figure}
\caption {Representative z$>$0.5 spirals in the
Hubble Deep Field whose images have sufficient signal/noise to
address questions concerning the star formation histories of
bulges, bars and stellar disks (c.f. Abraham et al 1999a,b;
Abraham \& Ellis 1999). Resolved stellar populations from such
multicolour images can be interpreted more clearly using dynamical
and metallicity information obtained from integral field
spectrographs.}
\label{fig-11}
\end{figure}

Whereas there is naturally much interest in surveying galaxies
with $z>$2-5, progress in understanding the astrophysical
processes which govern galaxy formation and evolution will also
require extensive studies of selected galaxies with the redshift
range 0$<z<$2 during which period it seems the Hubble sequence of
types became established. HST has shown the way with exquisite
optical imaging data. The challenge will be to complement this
with near-infrared and spectroscopic data of appropriate
resolution from ground-based telescopes and move away from the
traditional `number counting' approach which serves only to
determine the integrated properties of the galaxy population.

\section{Instrumental Themes}

The above scientific overview is intended to set the scene for
some of the 3-D applications discussed at this meeting. A few
common instrument themes are relevant in connecting the various
scientific questions:

\begin{itemize}

\item {\it Wide Field Imaging:} The Schmidt telescope revolution
which influenced this field in the 1970's has been replaced by its
digital equivalent - panoramic CCD cameras on 2.5-4m telescopes
with $\simeq$deg$^2$ fields. Already such surveys are being
extended into the near-IR. Multicolour surveys promise great
progress in the cosmological arena through supernova searching,
weak lensing and photometric redshifts aimed at tracking the
growth of galaxy correlations. There are substantial challenges in
each application, however. The Type Ia SN programme is hindered by
the need for more detailed astrophysical information spanning a
wide wavelength range (at both high and low z) rather than gross
statistics. Weak lensing as a probe of large scale structure and
$\Omega_M$ requires exquisite correction of instrumental effects
and may, ultimately, only be possible in space\footnote{At the
moment there is no contest as the field of view of Hubble Space
Telescope is very uncompetitive but a wide field modest aperture
space telescope would be overwhelmingly powerful in this area.}.
Photometric redshifts are largely untested in the all-important
1$<z<$2 range and redshift-dependent effects will have to be
understood before extracting meaningful signals on correlation
evolution.

\item {\it Wide Field Spectroscopy:} Now is also an exciting time for
wide field spectroscopic surveys. Although the prime spectroscopic
products of the Sloan and 2dF surveys will be maps of the $B<$20
galaxy distribution for power spectrum analyses, we can also hope
to utilise the galaxy spectra to investigate the astrophysical
origin of bias through luminosity or star-formation rate
dependencies. The spectroscopic surveys will deliver data that is
much richer than a simple redshift. Just as the Schmidt surveys
formed the basis for springboard projects using other facilities
so the 2dF survey is already proving useful for spinoff projects
from faint radio surveys to weak lensing behind known filamentary
structures. 2dF also has the potential to probe much fainter than
$B\simeq$20 and this raises the important question of an 8-m
equivalent device capable of exploring to $z\simeq$1 and beyond.

\item {\it Resolved Spectroscopy:} HST has shown us the
tantalising beauty of resolved galaxies at high redshift. At face
value it seems the Hubble sequence of galaxy types was laid down
by some process soon after redshift 2. Optical surveys are biased
however, being sensitive only to sources with recent or ongoing
star formation. We seek to extend our viewpoint of these galaxies
with high resolution images in the near-infrared, where more
representative stellar populations can be analysed, and through
far-infrared and sub-mm detections sensitive to obscured regions.
A major question is whether the optical/near-infrared should
remain the fulcrum of activity in unravelling galaxy evolution.
Our investment in this area is now unparalleled through HST and a
new generation of 8-m telescopes equipped for high resolution
work. Yet surprisingly little resolved spectroscopy has been done
at moderate look-back times. Together with upcoming facilities at
mid-IR and sub-mm wavelengths, we can expect to unravel the
details of galaxy evolution over 0$<z<$5 via studies of assembling
mass as well as the history of star formation.

\end{itemize}

\acknowledgments

I am grateful to my collaborators and students at Cambridge and
elsewhere for allowing me to present work done with their
assistance. I thank Joss Bland-Hawthorn and Wil van Breugel for
their encouragement and patience with this article.

\end{document}